# Basal-plane growth of cadmium arsenide by molecular beam epitaxy

David A. Kealhofer[1], Honggyu Kim[2], Timo Schumann[2], Manik Goyal[2], Luca Galletti[2], and Susanne Stemmer[2*]

[1]Department of Physics, University of California, Santa Barbara, CA 93106-9530, USA.

[2]Materials Department, University of California, Santa Barbara, CA 93106-5050, USA.

*Corresponding author.  Email: stemmer@mrl.ucsb.edu

**Abstract**

(001)-oriented thin films of the three-dimensional Dirac semimetal cadmium arsenide can realize a quantum spin Hall insulator and other kinds of topological physics, all within the flexible architecture of epitaxial heterostructures. Here, we report a method for growing (001) cadmium arsenide films using molecular beam epitaxy. The introduction of a thin indium arsenide wetting layer improves surface morphology and structural characteristics, as measured by x-ray diffraction and reflectivity, atomic force microscopy, and scanning transmission electron microscopy. The electron mobility of 50-nm-thick films is found to be 9300 cm$^2$/Vs at 2 K, comparable to the highest-quality films grown in the conventional (112) orientation. This work demonstrates a simple experimental framework for exploring topological phases that are predicted to exist in proximity to the three-dimensional Dirac semimetal phase.

The three-dimensional Dirac semimetal cadmium arsenide ($Cd_3As_2$) is distinguished among the expanding ranks of gapless topological materials by the fact that its two Dirac nodes – doubly degenerate band crossings along the $k_z$-direction that are protected by a symmetry of the crystal lattice – lie at the Fermi level and are isolated from other (parabolic) bands [1]. Its high electron mobility, large Fermi velocity, and chemical stability [2-8], make it attractive for future high-performance electronics. In epitaxial thin films, electrostatic gating, quantum confinement, and strain engineering afford separate, extra degrees of control over the electronic states, making such films an attractive platform for the study and realization of novel topological phases. The combination of $Cd_3As_2$'s unusual band structure with the tunability of thin film samples has led to the demonstration of the Dirac nature of the two-dimensional states [9] that give rise to the quantum Hall effect in very thin films [10-12].

To date, epitaxial thin film growth of $Cd_3As_2$, whether by molecular beam epitaxy (MBE) [13,14] or pulsed laser deposition [12,15], has been achieved on the tetragonal (112) plane, which is also the easy cleavage plane of single crystals [16]. The (001) plane differs from the (112) plane and indeed all others in that the two bulk nodes project onto one point, rather than two, of the surface Brillouin zone [1]. As the film thickness is reduced to the point when finite size effects become apparent (less than about 40 nm), the subbands alternate between an inverted and trivial ordering, and the material transitions between a quantum spin Hall insulator phase and a trivial band insulator phase [1]. The properties of films grown on the (001) plane are also likely to provide new perspectives on the nature of the quantum Hall effect and surface transport in (112) films. Furthermore, (001) oriented films may be required to observe Weyl fermions under applied magnetic fields [6].

Here we report on a MBE approach for the growth of reproducible, smooth (001) $Cd_3As_2$ films, with carrier mobilities comparable to the better-established (112) $Cd_3As_2$ films [10]. In particular, we demonstrate that the introduction of a thin InAs wetting layer significantly improves the structural characteristics. Signatures in quantum transport confirm the quality of these samples and suggest avenues for further research.

Samples were grown using conventional MBE techniques. A schematic of the heterostructure is shown in Fig. 1(a). After a solvent rinse, a cleaved piece of an undoped (001) GaSb wafer was loaded into a buffer chamber for outgassing. The chip was adhered to a tungsten plate with liquid gallium to improve temperature homogeneity and consistency in thermometry. The native oxide was thermally desorbed in the growth chamber under Sb flux, and a 150-nm-thick layer of GaSb was grown to smooth out any resulting pits before growth of a $Ga_xAl_{1-x}Sb_{1-y}As_y$ buffer layer for electrical isolation from the GaSb. While in principle an AlSb buffer layer electrically isolates the $Cd_3As_2$, the +0.649% lattice mismatch with GaSb can result in defects. The incorporation of small amount of Ga is thought to suppress the oxidation besides slightly reducing the lattice mismatch [17]; addition of As further reduces the lattice mismatch, allowing for the growth of thick buffer layers with low densities of extended defects. For example, high quality InAs quantum wells have been grown using $Ga_xAl_{1-x}Sb_{1-y}As_y$ buffer layers [18]. In quaternary alloys of Al, Ga, Sb, and As, the incorporation of the incident Al and Ga atoms is unity across a wide range of viable substrate temperatures, and as a result the desired group III composition can be achieved by setting the absolute flux of the sources. The relative Sb versus As incorporation, however, depends strongly on the substrate temperature. Here, the desired composition was achieved by maintaining a constant substrate temperature, measured by a pyrometer, and iterating on the beam equivalent pressure ratio of the group V sources. After the buffer layer, four monolayers (1.2 nm) of a thin wetting layer, either InAs, GaSb, or AlSb, were grown. The lattice constant of InAs differs from that of GaSb by –0.614%. We call the four monolayers a wetting layer because, given that the layer grows pseudomorphically on the underlying buffer layer, the differences in subsequent $Cd_3As_2$ growth on each must reflect differences in surface and/or interfacial energies. Finally, the growth parameters for (001) $Cd_3As_2$ were in most respects identical to those reported previously [13]. High flux from a molecular $Cd_3As_2$ source ($2.6 \times 10^{-6}$ torr on a flux monitoring ion gauge) was supplied to a low-temperature substrate (thermocouple temperature around 200 °C).

After structural characterization by atomic force microscopy (AFM) and x-ray diffraction (XRD) and reflectivity (XRR), Hall bar structures were fabricated using standard photolithographic techniques. Mesas were isolated by argon ion milling, and ohmic contacts were deposited as Au/Pt/Ti stacks. Unless stated otherwise, samples were exposed to a nitrogen plasma treatment prior to electrical measurement to reduce the density of low-mobility *p*-type carriers on the top surface [19]. Electrical measurements were performed in a Quantum Design Physical Properties Measurement System using standard lock-in techniques. Unless stated otherwise, the longitudinal magnetoresistance data shown was symmetrized and the Hall magnetoresistance antisymmetrized under a change in magnetic field (*B*) direction.

Cross-section transmission electron microscopy samples were prepared using a focused ion beam system with a final milling energy of 2 keV Ga ions. High-angle annular dark-field scanning transmission electron microscopy (HAADF-STEM) imaging was performed on an FEI Titan S/TEM ($C_s$ = 1.2 mm) operated at 300 kV with a convergence semiangle of 9.6 mrad. The experimental HAADF-STEM images were denoised by the Wiener filter [20]. The simulated image of Cd$_3$As$_2$ (space group I4$_1$/acd [16]) was calculated using the Kirkland multislice algorithm [21].

Figure 1(b) shows a high-resolution, out-of-plane 2θ-ω XRD scan of a sample with a 450-nm-thick Cd$_3$As$_2$ film on an InAs wetting layer and AlGaSb buffer layer. In the vicinity of the 002 GaSb substrate reflection (magnified in the left inset), the 008 Cd$_3$As$_2$ and 002 AlGaSb reflections are seen. The 1.2-nm-thick InAs wetting layer does not have enough volume to scatter x-rays above the detection limit. Another sequence of peaks is visible near the 004 GaSb reflection (center inset), starting with 00$\underline{16}$ Cd$_3$As$_2$. The kinematical scattering intensity of the 00$\underline{24}$ Cd$_3$As$_2$ reflection, expected near 006 GaSb (right inset), is less than that of the 008 Cd$_3$As$_2$ peak by a factor of more than $10^3$ and is not observed. The lack of other film peaks confirms the (001) orientation of the Cd$_3$As$_2$ film.

To verify the in-plane epitaxial alignment, the reciprocal space map in Fig. 1(c) shows a region in the vicinity of the 224 GaSb reflection. The diagonal line indicates the 224 peak position for a relaxed cubic layer. The vertical dashed line shows the position for a coherently strained film (i.e., in-plane lattice parameter identical to that of the substrate). The buffer layer peak falls between the two lines, which indicates partial strain relaxation, and its lattice parameter is about 0.5% larger than that of the substrate. From the position of the 44$\underline{16}$ Cd$_3$As$_2$ peak, its lattice parameters are determined to be $a$ = 12.65 Å and $c$ = 25.44 Å, which agree well with the published values for bulk Cd$_3$As$_2$ [16,22]. The $c/a$ ratio is 2.01, consistent with (001)-oriented Cd$_3$As$_2$ films.

The data shown in Figs. 1(d,e) are from samples grown under nominally identical conditions but on Ga$_x$Al$_{1-x}$Sb$_{1-y}$As$_y$ buffer layers with different wetting layers – InAs, AlSb, and GaSb, respectively. Figure 1(d) shows out-of-plane 2θ-ω XRD scans in the vicinity of the 004

GaSb reflection. The $Ga_xAl_{1-x}Sb_{1-y}As_y$ buffer layer composition can be calculated using $x = 0.8$ (from flux calibration) and regarding the alloy as a mixture of $Ga_{0.2}Al_{0.8}Sb$ and $Ga_{0.2}Al_{0.8}As$. Assuming a fully relaxed buffer layer, we find $y = 0.09$; in the case of a fully strained buffer layer, $y = 0.11$. This lattice mismatch relative to the $Cd_3As_2$ layer is − 4%. Despite the large mismatch, thin $Cd_3As_2$ films grown on heterostructures with InAs wetting layers show an elongation of the out-of-plane lattice parameter [cf. lines in Fig. 1(d)] that is, at least in part, due to residual epitaxial coherency strain (estimated to be about -0.7% in-plane). Furthermore, the differences in the scattered intensity of the 00$\underline{16}$ $Cd_3As_2$ peak reflects differences in their defect densities on the three wetting layers. The film on InAs is also smoother, because oscillations in XRR survive to higher scattering angles [see Fig. 1(e)] and this is also evident in the AFM images shown in Fig. 2. Nevertheless, the morphology of the (001) films differs from the smooth, step-like surfaces of (112) $Cd_3As_2$ MBE films [23]. Furthermore, as judged by the period of the fringes in XRR, $Cd_3As_2$ films grown on AlSb and GaSb are thicker than on InAs.

Figure 3(a) shows a cross-section HAADF-STEM image viewed along [100]. Shown are the $Cd_3As_2$/InAs interface and the top few unit cells of the buffer layer. Comparison with simulations [Fig. 3(b)], and especially the orientation of rows of subtly elongated butterfly-shaped features, which are caused by a displacements of cadmium atoms, further confirms the [001] growth direction, consistent with the lattice constants found in XRD.

Electrical transport measurements on three Hall bar devices of ~45-nm-thick $Cd_3As_2$ films on different wetting layers, measured at 2 K, are shown in Fig. 4(a). The similarity in the longitudinal magnetoresistance ($R_{xx}$) for films on AlSb and GaSb may be attributed to the granuality of the $Cd_3As_2$ films on both wetting layers (Fig. 2), which causes the current traveling through a small number of paths that are only weakly coupled to each other. By contrast, the device on the film with the InAs wetting layer [solid lines in Fig. 4(a)] shows a factor of 50 lower zero-field resistance. This is especially evident in Fig. 4(b), which shows only the results from the film on InAs. The incipient quantum Hall effect seen in Figs. 4(a,b) is a further sign of the high quality of the $Cd_3As_2$ film grown on InAs.

The differences in the longitudinal resistance between the three devices cannot be attributed solely to the difference in carrier concentration. Specifically, the 2D carrier concentrations, as determined by the Hall effect [Fig. 4(a)] are $1.0 \times 10^{12}$ cm$^{-2}$ (on InAs), $7.8 \times 10^{11}$ cm$^{-2}$ (on AlSb), and $6.5 \times 10^{11}$ cm$^{-2}$ (on GaSb), i.e. they vary by less than a factor of two. The differences in carrier concentrations correlate with the relative positions of the conduction bands in the three wetting layers [24], which suggests that band bending at the bottom interface may play a role. It is important to note, however, that transport is not through the wetting layer itself. One piece of evidence is the fifty-fold reduction in longitudinal resistance of samples with InAs wetting layers, relative to the other two, which is accompanied by only a two-fold increase in charge carriers. Furthermore, consider the raw (that is, not antisymmetrized or symmetrized) resistance data shown in Fig. 4(c) before and after a N* plasma treatment. In this sample, the $Cd_3As_2$ layer is slightly thicker (53 nm) and the carrier density is $1.3 \times 10^{12}$ cm$^{-2}$. As shown in an earlier study on (112) films [19], the low-energy nitrogen plasma changes the band bending at the top surface to result in high-mobility, single-carrier (*n*-type) 2D transport. As seen in Fig. 4(c), the condition of the top surface appreciably changes the measured resistance also for these (001) films. This is strong evidence that the measured resistance is not mainly due to a buried conductive layer. The sensitivity of the transport properties, including the emergent quantum Hall effect, to the surface treatment, hints at the importance of surface states also in (001) films. Finally, the Hall mobility of 9300 cm$^2$/Vs in the 45-nm-thick $Cd_3As_2$ film on InAs [Fig. 4(b)] is nearly identical to that of the thicker 53-nm-thick film, 9200 $cm^2$/Vs [Fig. 4(c)] and compares well to that reported for MBE-grown (112) films of similar thickness [10]. To sum up, (1) the sensitivity of the transport behavior to the condition of the top surface, (2) the behavior of the resistance relative to the carrier density across the three wetting layers, and (3) a similar resonse to the nitrogen plasma surface treatment, to what has been reflecting predominantely those of the $Cd_3As_2$ films.

In summary, we have demonstrated a method for growing high-quality, (001)-oriented epitaxial thin films of $Cd_3As_2$ on (001) III-V substrates. A thin InAs wetting layer improves the nucleation on (001) growth surfaces, resulting in smoother, coalesced films. An avenue of future studies are wetting layers with a lattice parameter matched to that of $Cd_3As_2$ to encourage layer by layer or step-flow growth, or at least further improve the surface morphology. Nevertheless, the carrier mobility demonstrates that these (001)-oriented $Cd_3As_2$ films can be viable counterparts to the more established (112)-plane films. The observed onset of a quantum Hall regime sets a clear path for future study of the nature of the two-dimensional states in this system and their relationship to the topological states as as controlled in a semiconductor-topological semimetal heterostructures.

**Acknowledgments**

The authors thank R. Kealhofer for help with the electrical measurements and A. McFadden and M. Pendharkar for suggestions regarding sample mounting and thermometry. The authors gratefully acknowledge support through the Vannevar Bush Faculty Fellowship program by the U.S. Department of Defense (Grant No. N00014-16-1-2814). Support was also provided by a grant from the U.S. Army Research Office (grant no. W911NF-16-1-0280). The microscopy work was supported by the U.S. Department of Energy (Grant No. DEFG02-02ER45994). This research made use of shared facilities of the UCSB MRSEC (NSF DMR 1720256).

**Figure Captions**

**Figure 1:** (a) Schematic of the heterostructure (not to scale). (b) Out-of-plane 2θ-ω XRD scans from a 450-nm-thick (001) $Cd_3As_2$ film grown on an AlGaSb buffer layer with a 1.2-nm-thick InAs wetting layer. The vertical lines show literature values for the GaSb substrate (solid) and $Cd_3As_2$ layer (dashed). The reflections around 00$l$ GaSb ($l$ = 2, 4, and 6 from left to right) are shown in more detail in the insets. The 00$l$ $Cd_3As_2$ reflections have $l$ = 8, 16, and 24, but the 00$\underline{24}$ $Cd_3As_2$ reflection is too faint to be seen. Unindexed peaks are from the AlGaSb buffer layer. (c) Reciprocal space map in the vicinity of the 224 GaSb reflection performed on the same film as shown in (b). The axes show the out-of-plane and in-plane components of the scattering vector, $q_{\parallel[001]}$ and $q_{\parallel[110]}$, respectively. The diagonal line (gray, solid) shows the cubic condition, i.e. $a = c$, or $q_{\parallel[110]} = (\sqrt{2})\, q_{\parallel[001]}$. The short vertical line (gray, dashed) shows the condition for the buffer layer being fully strained to the substrate. The $Cd_3As_2$ peak is not expected to fall along the cubic condition line because its lattice is tetragonal. (d) Out-of-plane 2θ-ω x-ray diffraction scan in the vicinity of 004 GaSb for ~45 nm films grown on different wetting layers, InAs, AlSb, and GaSb [see legend in (e)]. The buffer layer (second most intense peak) is a quaternary alloy $Ga_{0.2}Al_{0.8}Sb_{0.9}As_{0.1}$. The scans are offset for clarity. Two literature values for the 00$\underline{16}$ $Cd_3As_2$ scattering angle are indicated with vertical lines, one in agreement with the reflections in (b) [22] (dot-dash, teal), and the other corresponding to a slightly smaller lattice constant (dot, black) [16]. The fact that the 00$\underline{16}$ $Cd_3As_2$ peak in the top trace (InAs wetting layer) does not fall along the dot-dash line is likely due to residual epitaxial strain. (e) XRR of the three films shown in (d). The film grown on InAs is the smoothest and also apparently thinnest, judged by the tail-off of the oscillations and their period, respectively, even though all growth parameters were held constant.

**Figure 2:** AFM images of the surfaces of the three $Cd_3As_2$ films shown in Figs. 1(d,e), grown on different wetting layers (see labels). The lateral scan dimensions are 2 μm × 2 μm in (a-c). Note the different height scales, indicated in the bottom left corner of all images. Lower magnification images are shown in (d-f), with lateral dimensions are 20 μm × 20 μm. All images have undergone a first-order plane leveling procedure. The root-mean-square roughness values of the surfaces shown in panels (a), (b), and (c) are calculated to be 1.2 nm, 3.4 nm, and 3.0 nm, respectively.

**Figure 3:** (a) HAADF-STEM image of a $Cd_3As_2$ film on an InAs wetting layer with a $Ga_{0.2}Al_{0.8}Sb_{0.9}As_{0.1}$ buffer layer, recorded along [100]. (b) Magnified image of the $Cd_3As_2$ film, simulated image, and schematic. Due to the displacement of Cd atoms, the projected Cd atomic columns are seen as a characteristic elongated, butterfly-like shape (purple shaded ellipses), which form rows parallel to [010].

**Figure 4:** (a) Comparison of 2 K magnetotransport behavior of $Cd_3As_2$ films differing only by wetting layer. These are the same films as in Figs. 1(d,e) and Fig. 2. The longitudinal resistance ($R_{xx}$) is shown in shades of blue and the Hall resistance ($R_{xy}$) is shown in shades of red. The device sizes are: 10 μm × 10 μm on InAs (solid), 50 μm × 50 μm on AlSb (dot-dash), and 100 μm × 100 μm on GaSb (dash). (b) Higher magnification of magnetoresistance behavior for the 45-nm-thick $Cd_3As_2$ film on InAs wetting layer shown in (a). Strong quantum oscillations in the longitudinal magnetoresistance (solid blue) are accompanied by incipient plateaus in the Hall magnetoresistance (red dashes). (c) Comparison of magnetoresistance behavior for a 53-nm-

thick $Cd_3As_2$ film on an InAs wetting layer before (dashes) and after (solid) surface treatment by a low-energy nitrogen plasma. The more symmetric $R_{xx}$ behavior after cleaning (dark blue) and more antisymmetric $R_{xy}$ (orange) after cleaning reflect the fact that the charge transport is less scattered and thus there is less mixing between the two channels. Unlike data in (a,b), these data were not symmetrized or antisymmetrized with respect to reversal of the magnetic field.

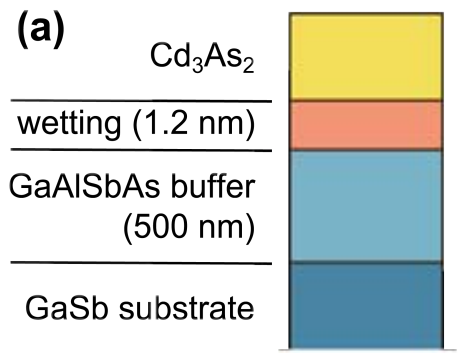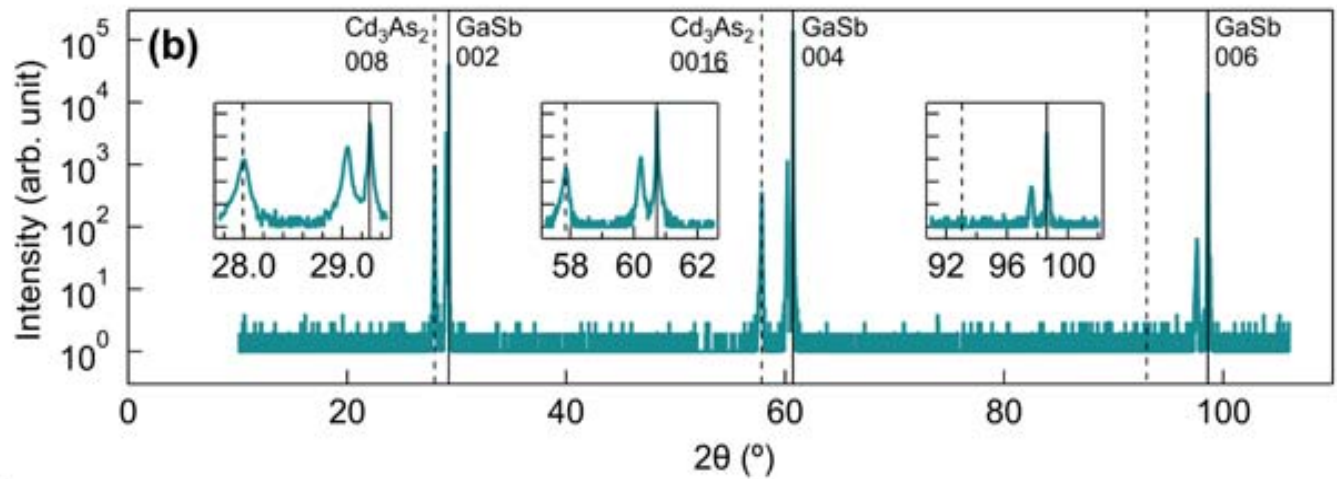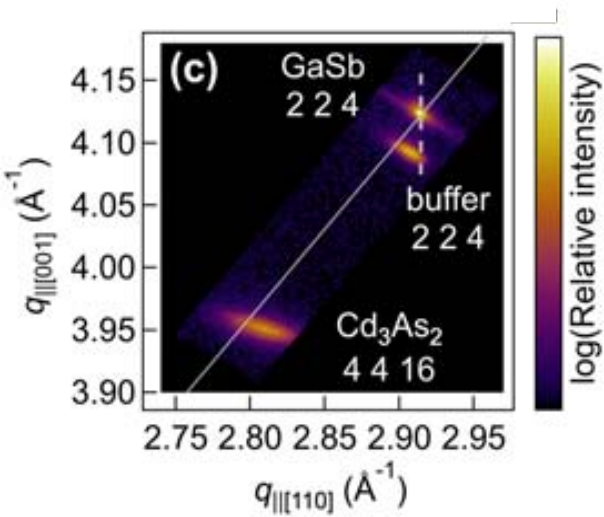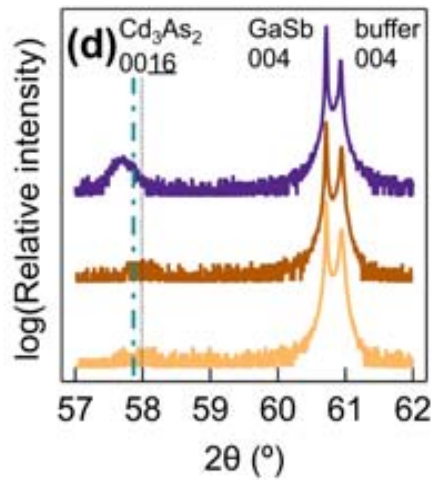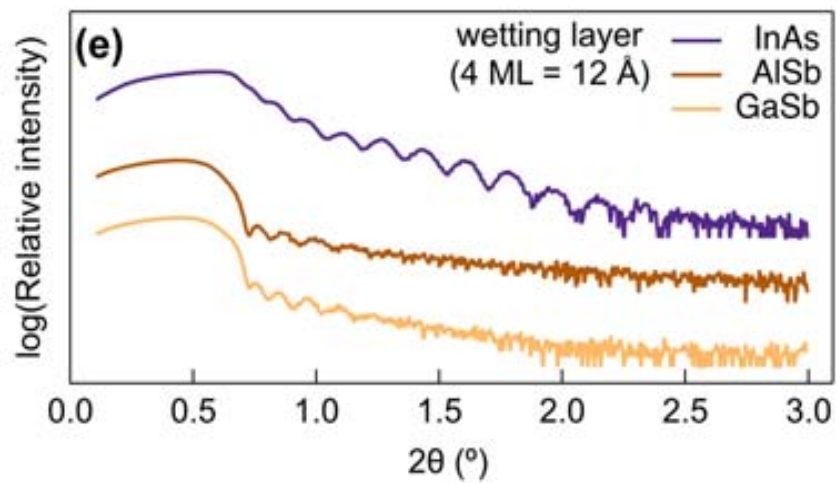

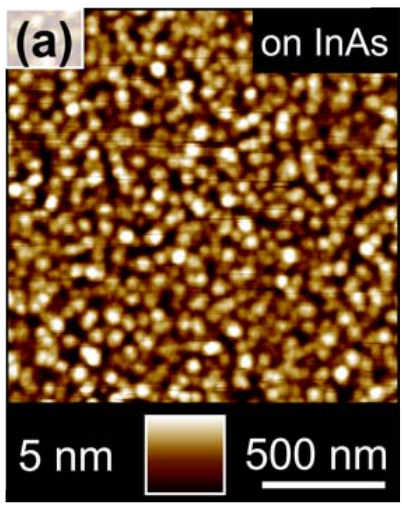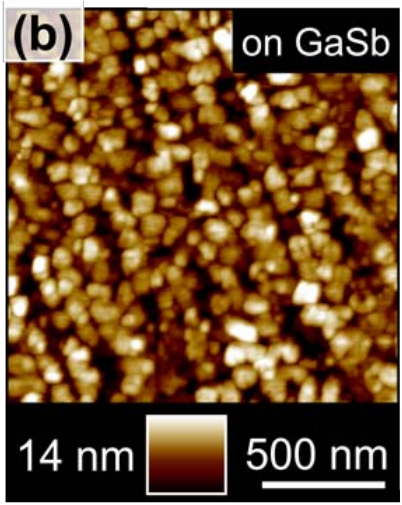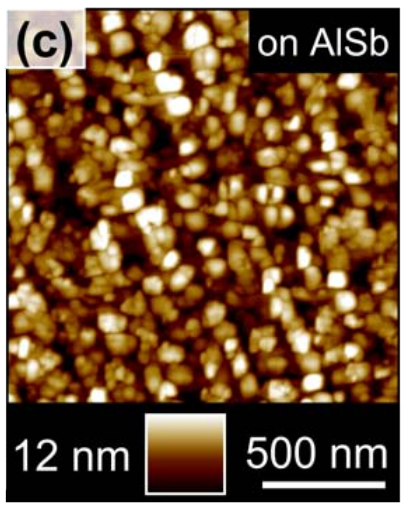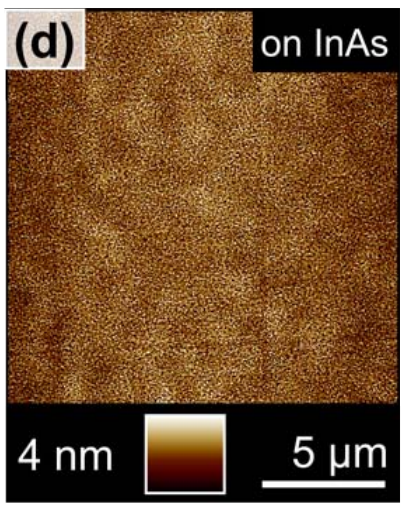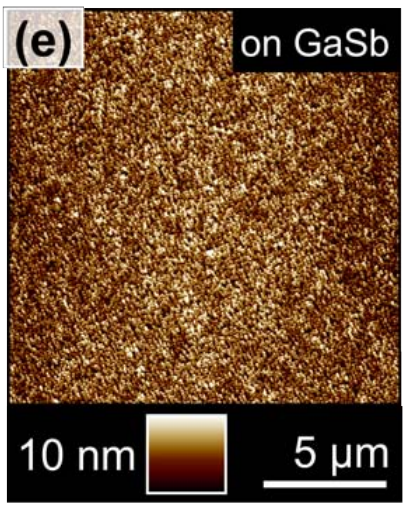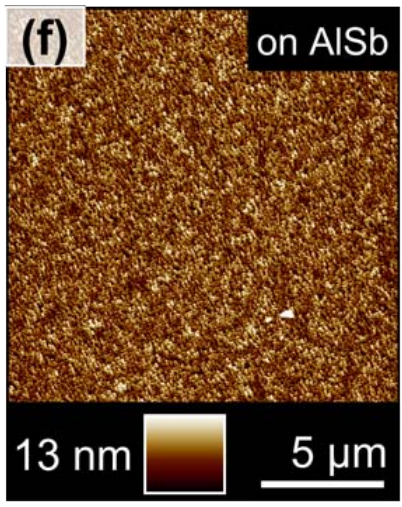

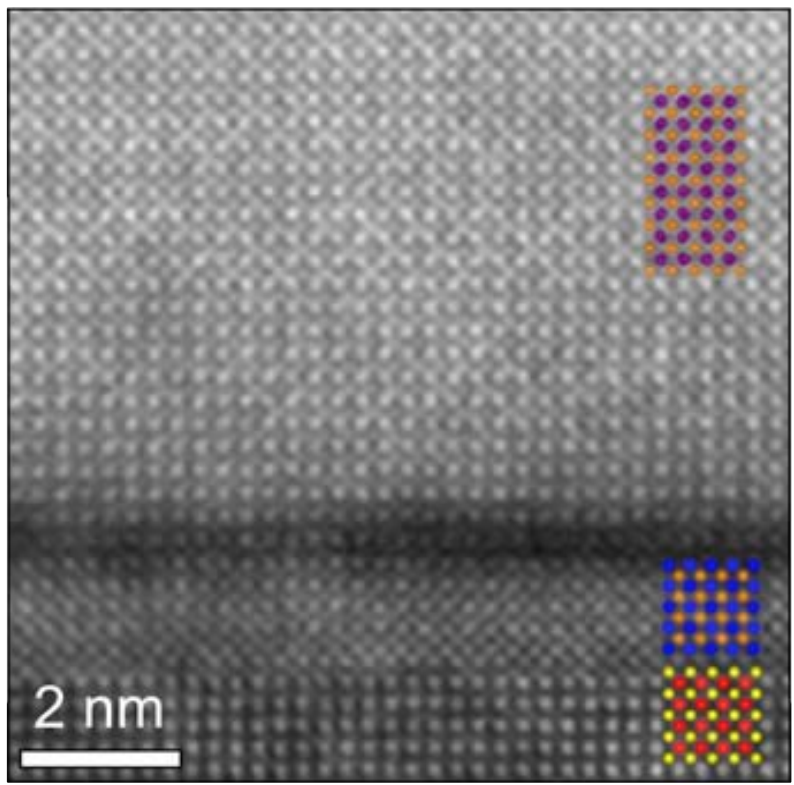

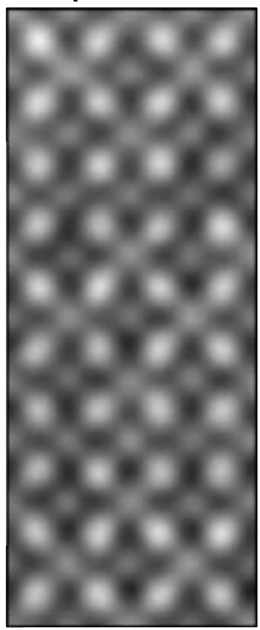 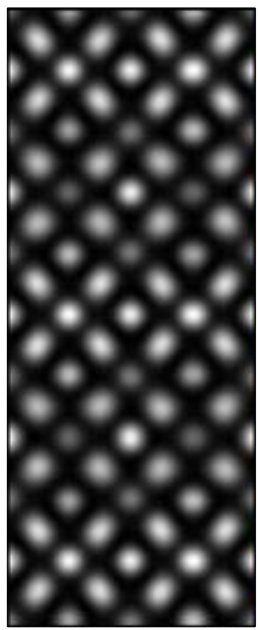 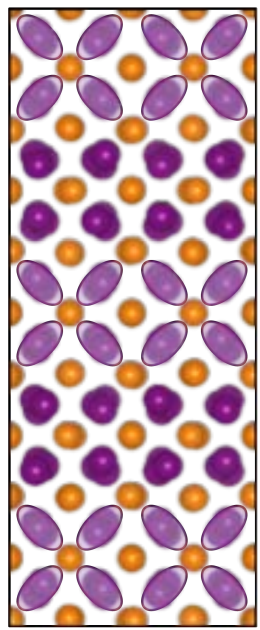

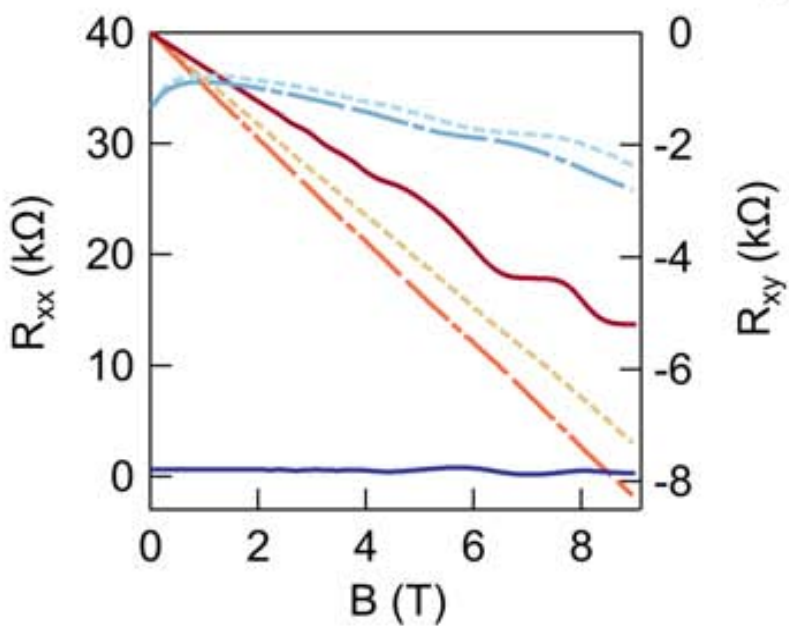

(a)

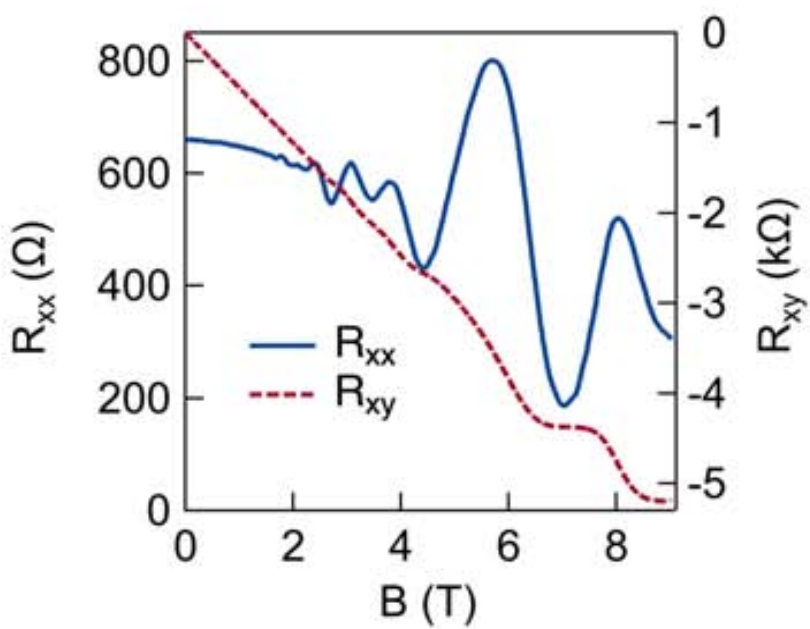

(b)

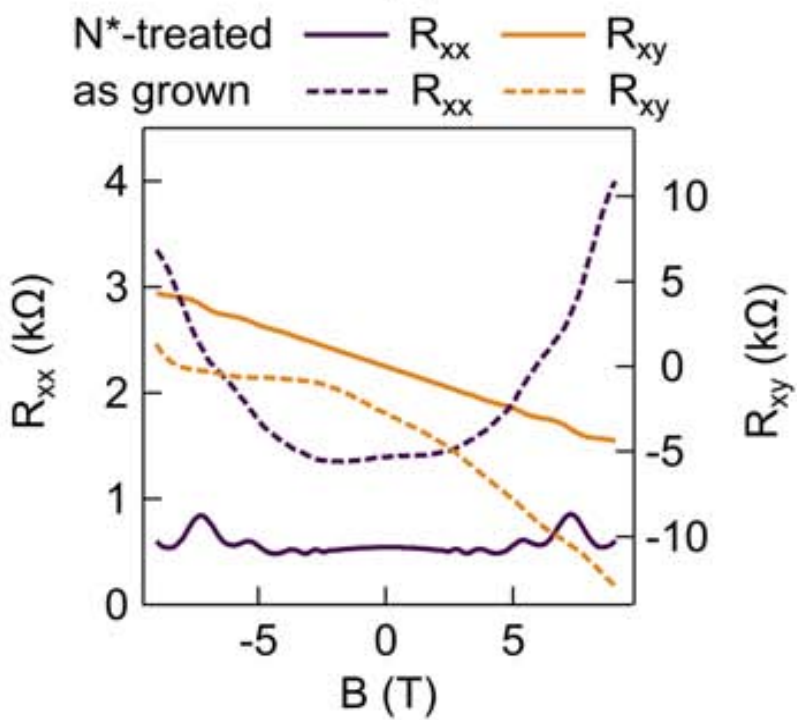

(c)